\documentclass[12pt,preprint]{aastex}

\newcommand{\simgt}{\ga}
\newcommand{\simlt}{\lower.5ex\hbox{$\; \buildrel < \over \sim \;$}}

\slugcomment{draft date: \today}

\begin{document}

\title{The Dust cloud around the White Dwarf G 29-38}

\author{William T. Reach\altaffilmark{1},
Marc J. Kuchner\altaffilmark{2},
Ted von Hippel\altaffilmark{3},
Adam Burrows\altaffilmark{4},
Fergal Mullally\altaffilmark{3},
Mukremin Kilic\altaffilmark{3},
D. E. Winget\altaffilmark{3}
}

\altaffiltext{1}{{\it Spitzer} Science Center, MS 220-6, 
California Institute of Technology,
Pasadena, CA 91125}

\altaffiltext{2}{NASA Goddard Space Flight Center, Greenbelt, MD 20771}

\altaffiltext{3}{Department of Astronomy, University of Texas, 1 University Station C1400, Austin, TX 78712}

\altaffiltext{4}{Department of Astronomy and Steward Observatory, University of Arizona, 933 North Cherry Avenue, Tucson, AZ 85721}

\email{reach@ipac.caltech.edu}

\begin{abstract}

We present new observations of the white dwarf G 29-38 with
the camera (4.5 and 8 $\mu$m), photometer (24 $\mu$m),
and spectrograph (5.5--14 $\mu$m) 
of the {\it Spitzer} Space Telescope.  This star has an exceptionally
large infrared excess amounting to 3\% of the bolometric luminosity.
The spectral energy distribution has a continuum
peak around 4.5 $\mu$m and a 9--11 $\mu$m emission feature
1.25 times brighter than the continuum. 
A mixture of amorphous olivine and a small amount of forsterite
in an emitting region 1--5 $R_\odot$ from the star can
reproduce the shape of the 9--11 $\mu$m
feature.  The spectral energy distribution also appears to require
amorphous carbon to explain the hot continuum.
Our new measurements support the idea that 
a relatively recent disruption of a comet or asteroid created the cloud.

\end{abstract}


\section{Introduction}

White dwarfs offer a unique view into the properties of planetary systems.
A star can retain much of its planetary system as it leaves the main sequence
and evolves  through the red giant stage. 
Although the inner $\sim 5$ AU of a planetary system may 
evaporate within the red giant atmosphere and planets
that were in marginally stable orbits may escape during stellar mass loss,
most outer planets and even much of
the Oort-cloud-like extremities of planetary systems should persist
around white dwarfs \citep{debes02}.

The white dwarf Giclas 29-38 (WD 2326+049; G29-38 hereafter) 
has garnered intense interest since \citet{ZB87}
discovered infrared emission from this object far in excess of the
photosphere.   The excess emission,  confirmed
both by ground-based \citep{tokunaga} and space-based \citep{chary} observations,
begins to deviate from a pure photosphere
in the near-infrared.   While it was initially speculated that a Jupiter-sized companion could
supply all the infrared excess, follow-up observations
(e.g. Kuchner et al. 1998, Graham et al. 1990)  appear to
have ruled this out.  The colors measured with ISOCAM 
suggested the infrared excess is due to particulate matter rather than 
a brown dwarf companion \citep{chary}. 

Besides strong hydrogen absorption lines,
G29-38 has some photospheric metal lines, including \ion{Ca}{2}
which should rapidly ($< 10^4$ yr) sink out of the atmosphere due to
gravitational sedimentation
\citep{alcock80,koester97,zuckreid}.
Possible explanations for the atmospheric metals 
and the infrared excess include interstellar medium (ISM) accretion 
\citep{dupuis,koester97,zuckreid}, 
cometary breakup \citep{alcock}, and asteroid disruption \citep{jura}.


We turned the powerful instruments of the 
{\it Spitzer} Space Telescope \citep{werner} towards G29-38 to
measure its brightness from 4.5--24 $\mu$m.  These observations probably
probe a planetary system in a late stage of evolution.
G29-38's main sequence progenitor was probably an A star with mass
approximately 3.1 $M_\odot$ \citep{weidemann}, similar to many stars
known to have debris disks and potential planetary systems \citep{riekeAstar}.
The post-AGB phase age is estimated to be $\sim 5\times 10^8$ yr
\citep{postAGBage}.


\def\extra{
infrared excess 2--5 $\mu$m, far above photosphere, Tcolor=1200 K,
title: "an orbiting white dwarf?" 
tauPR for particle radius 0.5 micron --> 10 yr
minimum mass of grains 3e17 g
"accumulating this much orbital material in such close proximity
to G29-38 seems very unlikely given the rapid depletion due to the Poynting-Robertson effect
and the absence of any spectral peculiarities in its photospheric spectrum"
adopt distance 14 pc \citep{ZB87}
}

\begin{figure}[th]
\plotone{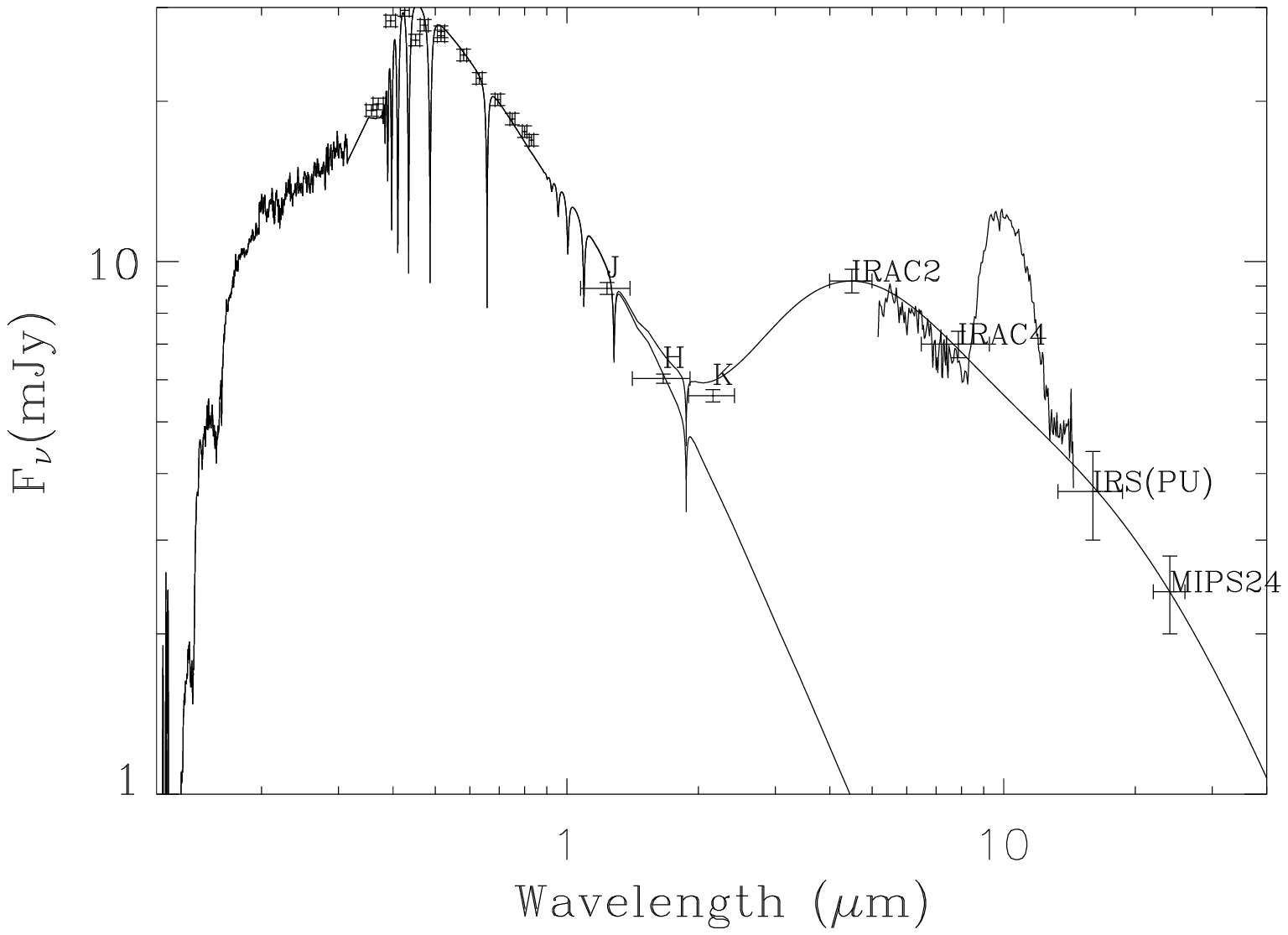}
\caption[f1.eps]{
Spectral energy distribution of the white dwarf G29-38
from the ultraviolet through infrared. The infrared observations
are described in \S2 and the ultraviolet through
near-infrared observations and model are described in \S3.
The solid line through the infrared data is a modified blackbody
fit to the continuum.
\label{wdspec}}
\end{figure}

\section{Observations\label{obssec}}

Figure~\ref{wdspec} shows the spectral energy distribution of
G29-38. 
Observations with the Infrared Array Camera (IRAC) \citep{fazio}
were performed on 2004 Nov 26 10:54 UT. 
At each of five dithers 
in a Gaussian spatial distribution, a 30 sec frame was taken
with the 4.5 and 8 $\mu$m arrays.
We extracted our photometry using the method \citep{reachcal} used 
for the IRAC calibration stars.  The quoted 4.5 and 8 $\mu$m fluxes, measured
from the basic calibrated data,
are $8.8\pm 0.3$ and $8.2\pm 0.3$ mJy, respectively.
We calculated color corrections of
0.995 and 1.17 in the 4.5 and 8 $\mu$m channels, making the estimated
fluxes 9.2 and 7.0 mJy at the IRAC nominal
wavelengths of 4.493 and 7.782 $\mu$m.
Observations with the Multiband Imaging Photometer for Spitzer
(MIPS) \citep{rieke} were performed on 2004 Dec 2 04:45 UT. 
Three cycles of small-scale photometry dithers were taken
with 10 sec frames (total exposure time 420 sec); the 
flux at 24 $\mu$m is 2.4 mJy.

Observations with the Infrared Spectrograph (IRS) \citep{houck}
were performed on 2004 Dec 8 3:26 UT. 
The source was first observed on the blue-filter portion of the
peak-up array;
the flux measured from the 16 $\mu$m peak-up image is 3.7 mJy.
The spectrum from 5.2 to 14.4 $\mu$m was assembled by differencing
observations at two nods on each of the 4 spectral orders.
The flux derived from the 
IRS spectrum convolved with the IRAC 8 $\mu$m bandpass
is consistent with the IRAC flux.

The IRS spectrum shows continuum emission and a strong 9--11 $\mu$m
feature characteristic of silicate minerals. It shows none of the
bands at 6.2, 7.7, 8.6, 11.3, or 12.6 $\mu$m characteristic of
polycyclic aromatic hydrocarbons that normally dominate
mid-infrared spectra of the ISM.
The excess emission above photospheric is approximated by
two modified blackbodies with temperatures 890 K and 290 K
and dilution factors (at 10 $\mu$m wavelength)
$4.2\times 10^{-16}$ and $6.3\times 10^{-15}$,
and emissivity proportional to $\nu^{0.5}$.
These functions are only intended as mathematical
approximations of the continuum shape.
\def\extra{
If the emission were optically thick
at 10 $\mu$m, the dilution factors would imply,
for the two temperature components, emitting regions
of size 5 and 18 mas, and distances from the star
of 7 and 27 $R_\odot$, respectively. These are lower
limits to the size of the emitting region, which in fact 
must be optically thin at 10 $\mu$m in order to present
such a high-contrast silicate emission feature. 
}

\section{Modeling\label{modsec}}

We adopt the
following basic parameters for G29-38. It is a
DA4 type white dwarf, with atmospheric spectrum 
dominated by H lines \citep{green}. 
It has an assumed mass of 0.69 $M_\odot$,
surface gravity $\log g=8.14$, and  
temperature $11,800$ K \citep{bergeron95}.
The parallax is
$\pi=0.071''+/-0.004''$ \citep{usno}, implying a distance of 14 pc.

We assembled the spectral energy distribution
from the IUE low-dispersion spectrum 
covering 0.115--0.3148 $\mu$m \citep{holberg}
and a model atmosphere for $T=12,000$ K
and $\log g=8$ covering 0.35--60 $\mu$m (courtesy D. Koester).
We used the 2MASS photometry ($J=13.132\pm 0.029$, $H=13.075\pm 0.022$, 
$K_s=12.689\pm 0.029$) and optical spectrophotometry 
from Palomar \citep{greenstein} for normalization, giving
extra weight to the 2MASS J-band photometry,
the longest-wavelength where we think the photosphere dominates.
The integrated luminosity of the star is $2\times 10^{-3} L_\odot$.

\begin{figure}[th]
\epsscale{.80}
\plotone{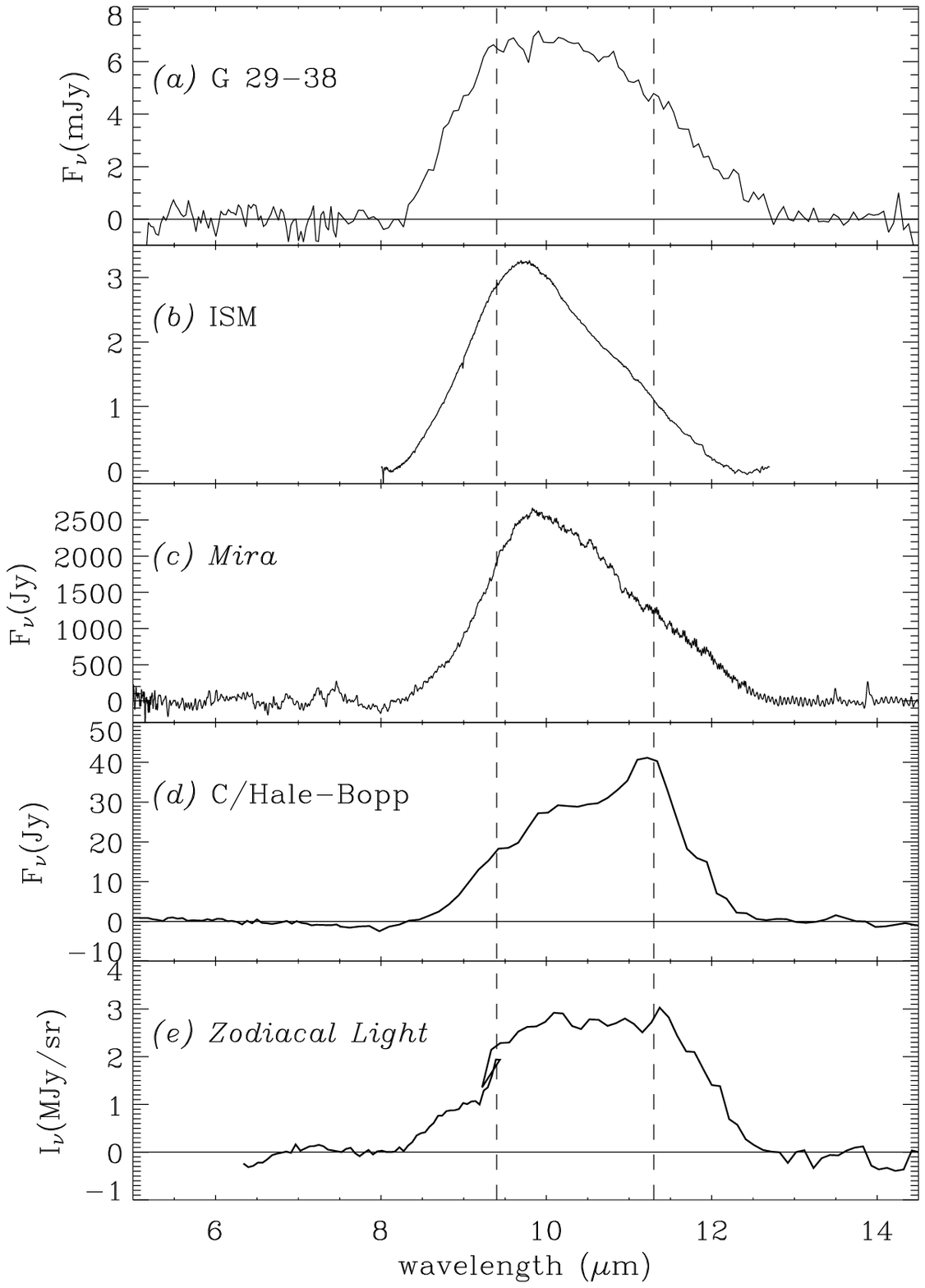}
\caption[f2.eps]{
Continuum-subtracted spectrum of the silicate feature from G29-39,
compared to similarly subtracted silicate features of the diffuse
ISM, the O-rich star Mira, 
comet Hale-Bopp, and the zodiacal light.
\label{wdspec_sil}}
\end{figure}

The 9--11 $\mu$m emission feature indicates silicates. 
Figure~\ref{wdspec_sil} compares the shape of the G29-38 silicate feature
to that of interstellar dust \citep{kemper},
Solar System zodiacal light, the O-rich mass-loss from the star Mira \citep{sloan},
and comet Hale-Bopp \citep{crovisier}.
The G29-38 silicate feature is redder than that of the ISM,
which has a single, rounded peak at 9.7 $\mu$m and nearly linear slopes 
on either side. This suggests that the material around G29-38 is not
accreted interstellar matter but rather is indigenous to the star.
The G29-38 silicate feature is more compact than the zodiacal light
silicate feature as well as the exozodiacal light around $\beta$ Pic feature \citep{reachzody}.
The Hale-Bopp spectrum has a prominent 
11.3 $\mu$m peak that is not seen in the G29-38 spectrum.
The Mira silicate feature is somewhat redder and wider than the ISM feature, roughly
intermediate between the ISM and G29-38 feature shape.
The G29-38 silicate feature shape is most similar to (and approximately
intermediate between) those of Mira and the zodiacal light, but the line-to-continuum
ratio of G29-38 is much greater than that of the zodiacal light (125\% versus 6\%)
suggesting a very different particle size distribution. 

We computed theoretical emission spectra for grains
(size $a=$0.01--1000 $\mu$m, $dn/da\propto a^{-3.5}$)
of various compositions (amorphous carbon, crystalline enstatite,
amorphous pyroxene [50/50 Mg/Fe], amorphous
olivine [50/50] and crystalline forsterite)
around G29-38 to further elucidate the properties of
the dust around that star.
The temperatures and emissivities over a
range of distances (0.003--3.5 AU = 0.6--750 $R_\odot$) 
from the white dwarf were integrated over a spherical,
presumed-optically-thin shell with a radial profile
$n\propto r^{-\alpha}$ and an inner cutoff at $R_{min}$.
The temperatures of 0.5 $\mu$m radius grains of
amorphous carbon (olivine) are 
600 K (920 K) at 3 $R_\odot$ and
130 K (110 K) at 1 AU from the white dwarf.
\def\extra{
The temperature 
can be approximated by $T=T_1 r^{-\delta}$, with $\delta\sim 0.4$--0.5
and $T_1=212$ K for small carbon grains, 65 K
for large olivine or pyroxene grains, and 33 K for pure crystalline
enstatite. 
Comparing to the observed spectral shape, we require the 
region emitting at 10 $\mu$m to be $<10 R_\odot$, and 
the region emitting at 24 $\mu$m to be $<100 R_\odot$.
}
Grains hotter than 1500 K are presumed sublimated and are
not included in the emissivity.
\def\extra{
The observed color temperature (from the continuum shape
near 10 $\mu$m wavelength)
is $> 700$ K, which requires that the bulk of the observed emission
arises within $10 R_\odot$ of the star, even for submicron grains.
The 24 $\mu$m color temperature $\sim 290$ K can be produced
by small grains at 30--100 $R_\odot$ or large grains at 9 $R_\odot$
from the white dwarf. 
The emitting region sizes inferred from the color
temperatures are consistent with the lower limits from the 
dilution factors. Thus we shall model the emission from a cloud
extending from an inner radius of 0.2 $R_\odot$ to an outer
radius of 1 AU, and consider the compositions and radial density
profile consistent with the observed emission.   WHAT DOES THIS MEAN?
}
Since we assume the cloud is optically thin, we cannot further
constrain the geometry of the emitting region, which could
range from a spherical shell to a flattened disk.

The G29-38 silicate feature resembles the
amorphous olivine model, with excess on the red side
that can be accounted for by mixing in forsterite.
Pyroxene can not contribute significantly to the observed spectrum;
its 9--11 $\mu$m feature is too blue. 
Amorphous olivine and fosterite alone cannot explain the spectrum,
however, because it cannot emit at 3--6 $\mu$m
without also producing a 1.6 $\mu$m feature that exceeds
the near-infrared H-band flux.
The near-infrared continuum shape at these
wavelengths matches the amorphous carbon model.
The contrast of the silicate feature demands a
particle size distribution that favors small particles, unlike that of
interplanetary dust grains \citep{gruen85}.
On the assumption that all the particles participate in the same
collisional evolution, we assume that silicate and carbon particles
share the same size distribution.  


\begin{figure}[th]
\epsscale{1}
\plotone{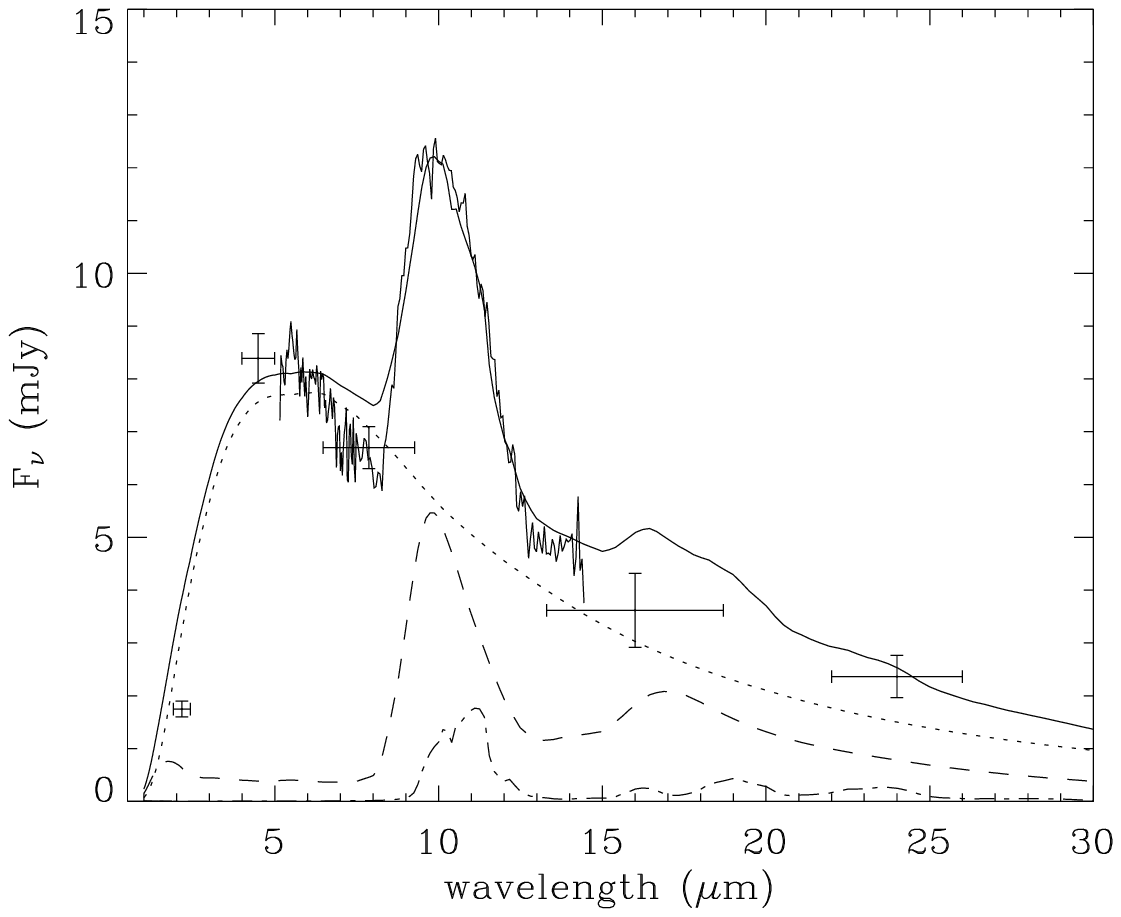}
\caption[f3.eps]{The infrared excess of G29-38 is compared to
the best-fit model (smooth solid line) consisting of three compositions: 
amorphous carbon (dotted line), amorphous olivine (long-dashed line),
and forsterite (short-long dashed line).
\label{fplot}}
\end{figure}

The distance of the emitting region from the star is constrained as follows.
In order to produce significant
emission at 5 $\mu$m, we require $R_{min}<5 R_\odot$. 
The best-fitting models all require small $R_{min}$, implying 
that the dust exists nearly up to its sublimation temperature.
The radial profile shape requires high $\alpha$, suggesting the 
emitting region extends from 1--10 $R_\odot$ from the star.
Figure~\ref{fplot} shows the observed spectrum and the best-fit model.
Experiments with other geometries yielded similar results:
a face-on slab required $\alpha\sim 3$, $R_{min}\sim 2.5 R_\odot$.
A dust cloud this small is consistent with the measurements of 
\citet{kuchner98} that showed G29-38 to be unresolved in K-band
at 55 mas  (160 $R_\odot$) resolution.

\section{Conclusions}

A cloud of small grains 1--10 $R_\odot$ from G29-38 creates its
mid-infrared emission. 
The luminosity of the infrared excess is 3\% of the luminosity of the
star, high by the standards of debris disks, indicating nontrivial
disk opacity in the ultraviolet-visible range.

Our models suggest a dust grain abundance ratio (by number) 
of olivine:carbon:forsterite of 5:12:2. 
Considering only the atoms in these grains, 
assuming material densities of 2.2 and 2.5 g~cm$^{-3}$
for the amorphous carbon and silicates, 
the abundance ratio (by number) of C:O is 3:1. 
The total dust mass required to generate the observed mid-infrared
flux is of order $10^{18}$ g; a larger mass could be present if
there are larger, cooler grains that do not contribute to the observed flux.
This inferred mass corresponds to that of a $\sim$10 km
diameter asteroid or comet, slightly less than
that of the interplanetary dust responsible 
for the zodiacal light in the Solar System \citep{fixsendwek}.


Where could this dust cloud have come from? It was not 
generated by the white dwarf (by any known mechanism), nor
is it likely to be planetary system material that was at
this location before the star was a red giant, whose
atmosphere extended much further than the current dust cloud.
The silicate feature is suggestive of dust formed in O-rich
mass-loss during the red giant or AGB phase, but the presence of 
significant and comparable amounts of silicate and carbonaceous dust 
(and no PAH) does not seem compatible with such dust, which 
in any event could not have survived at 10 $R_\odot$ from the star.
The material must have been transported inward.
If the material originates from the planetary system of the progenitor,
then gravitational perturbations would occasionally send smaller
bodies close to the star.
If the small body is collisionally fragmented, or loses mass
by cometary sublimation, or breaks apart by thermal or gravitational stress,
the Poynting-Robertson effect will cause particles of radius $a$ ($\mu$m) and
distance $r$ ($R_\odot$) to spiral into the star on timescales
of $4 r^2 a$ yr, i.e. a few years for
material contributing to the mid-infrared emission.
The particles would continue toward
the star until they sublimate, suffer mutual collisions,
disrupt, with part of the material landing in the
photosphere, part being blown out of the 
system by radiation pressure, and part remaining on bound orbits. 
If the entire mid-infrared-emitting
mass (i.e. particles out to $\sim 10 R_\odot$)
were lost in the Poynting-Robertson timescale, the accretion
rate is of order $10^{15}$ g~yr$^{-1}$. 
This flow could be supplied by collisional comminution of debris from
an asteroidal and cometary cloud like the one in the Solar System
\cite{sykes86}.  
This mass loss rate is comparable to the `high' ISM accretion
scenario calculated by \citep{dupuis}, which appears inconsistent
with G29-38's present location in a low density ISM environment.

A plausible model for the infrared excess around G29-38 involves
the tidal disruption of an asteroid near the white dwarf \citep{graham,jura}.
Since short-period comets are composed largely of refractory material,
we generalize this model to include the tidal disruption
of a comet.  Comets are commonly observed to
pass near the Sun, so by analogy a surviving reservoir of comets
(and possibly a surviving massive outer planet) would inevitably
lead to star-grazing cometary passages. With its low
luminosity, the white dwarf would not drive a high sublimation
rate for star-grazing comets, reducing the related stresses that appear to
cause Sun-grazing comets to split. Comets may then travel rather
close to the white dwarf, where they could be disrupted by
tidal forces.  Our new observations support this model, since the
emitting region lies close to the star, where a large solid body
would be tidally disrupted. 
Based on the high C/O ratio in the dust, it is also conceivable that
the `comet' was condensed out of the AGB mass-loss \citep{kuchnersaeger}.
The details of the tidal disruption model presented here are
different from previous papers which assume larger grains and
an optically thick disk. 
If the emitting region is a disk, it is either optically thin
or it is viewed relatively face-on, with the line of sight
from the white dwarf to the Sun relatively free of dust. 
For a thin disk to absorb 3\% of the luminosity of the
star it would have to be opaque in the ultraviolet,
where there is no evidence of significant extinction and no
2175 \AA~ absorption bump in the IUE spectrum.

An alternate and long-standing model for both the photospheric metals and the
infrared excess is dredge-up and mass loss.
There are already difficulties with dredge-up models associated with the need
for vertical velocities that are at odds with G29-38's g-mode pulsations 
that are overwhelmingly horizontal.  Further, standard white dwarf models are layered
from prior nuclear burning; 
hypothetical mixing from the pulsations would have to reach many orders of
magnitude below the convection zone boundary, well below even the degeneracy
boundary (about $10^{-6} M_{\odot}$) to dredge up material from the C/O layer
(presumably around $10^{-3}M_{\odot}$ or deeper).  
Further, the dredge-up hypothesis cannot explain
the abundant silicates; 
invoking these materials just below the He-layer is
very difficult to accommodate within our current understanding of nuclear
burning of He in post-main sequence stars appropriate to white dwarf stars in
the mass range of G29-38.  
Our observations therefore argue that the observed
material was not formed in the white dwarf; it formed well outside its current
position.

The presence of a dust cloud around G29-38 is likely related to
the anomalous presence of metals in its photosphere.
As discussed by \citet{zuckerman}, the photospheric metals could 
be accreted from interstellar gas, but then the
star would need to be within a dense cloud (because
the lifetime of metals in the photosphere [$< 10^4$ yr]
is smaller than the crossing time of a dense cloud [$\simgt 10^4$ yr]),
which is not true for G29-38.
The other major model for white dwarf photospheric metals is
cometary impact \citep{zuckreid,debes02}. Based on the presence of
a strong silicate feature and the overall mid-infrared 
spectral energy distribution, we believe this latter model
prevails for G29-38. 
The abundance of photospheric metals for G29-38
is among the highest of any white dwarf, and the infrared excess
is also the highest of any known white dwarf.
Another white dwarf, WD 1337+705 (G 238-44), with even higher 
atmospheric Ca abundance \citep{zuckerman} does not show infrared
excess at the level seen in G29-38
in our {\it Spitzer} photometric survey of white dwarfs 
at 4.5 and 8 $\mu$m.
It is possible that a recent disruption of an asteroid or comet has 
occurred in G29-38, re-populating the star's dust cloud
at a level much higher than the long-term steady state, as indeed
occurs in debris disk around main sequence stars \citep{riekeAstar} and
the Solar System \citep{sykes86}.

Are there planets around G39-38?
The 9--11 $\mu$m spectral feature proves there is a cloud of small 
silicate grains.  We ascribed the 3--6 $\mu$m continuum to amorphous carbon dust,
and the carbon-silicate mix matches the spectral energy distribution,
but there is no spectral signature for carbon other than its color temperature.
If we ascribe the 890 K continuum to a planet, then it must be 
not only very hot but also very large, with radius 
0.2 $R_\odot$---an extremely unusual object.
If the dust cloud is indeed due to an asteroid or comet, perhaps
other, cooler planets await discovery around
the fascinating star G29-38.

\acknowledgements  

This work is based in part on observations made with the {\it Spitzer Space
Telescope}, which is operated by the Jet Propulsion Laboratory, California
Institute of Technology under NASA contract 1407. Support for this work was
provided by NASA through award Project NBR: 1269551
issued by JPL/Caltech to the University of Texas.


\def\extra{

\clearpage

\begin{figure}[th]
\plotone{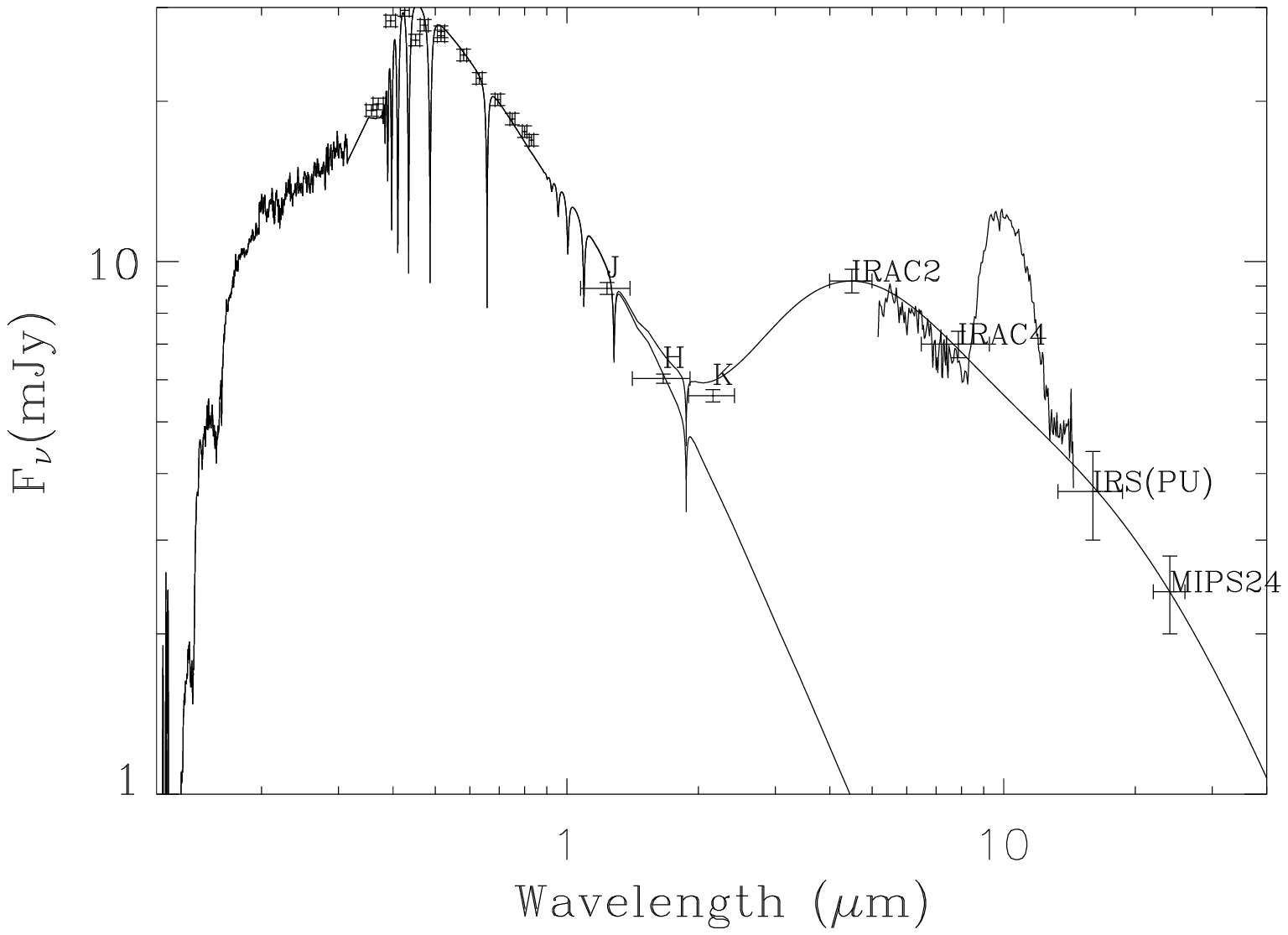}
\caption[f1.eps]{
Spectral energy distribution of the white dwarf G29-38
from the ultraviolet through infrared. The infrared observations
are described in \S\ref{obssec} and the ultraviolet through
near-infrared observations and model are described in \S\ref{modsec}.
The solid line through the infrared data is a modified blackbody
fit to the continuum.
\label{wdspec}}
\end{figure}

\clearpage

\begin{figure}[th]
\epsscale{.80}
\plotone{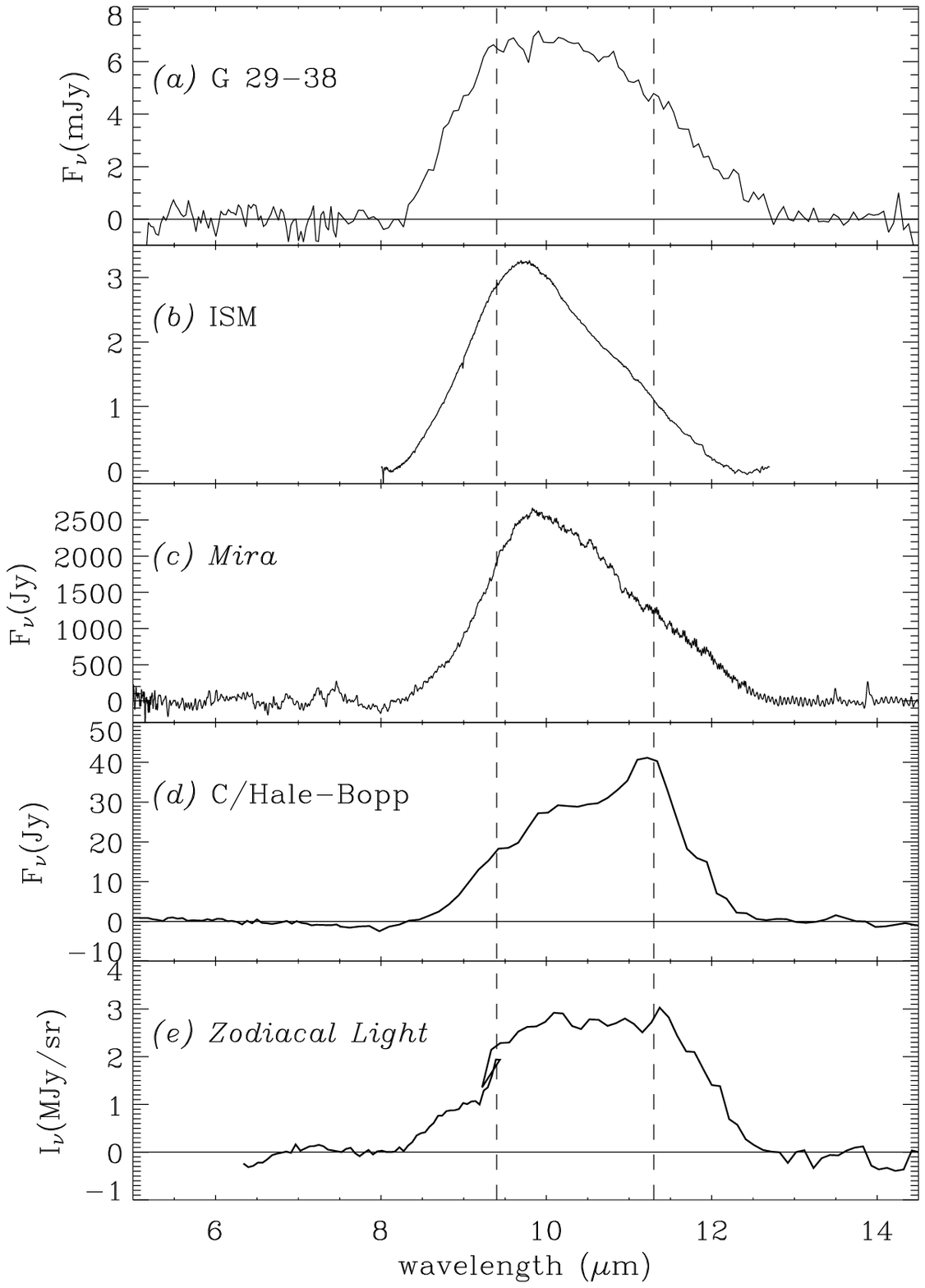}
\caption[f2.eps]{
Continuum-subtracted spectrum of the silicate feature from G29-39,
compared to similarly subtracted silicate features of the diffuse
ISM, exozodiacal light around $\beta$ Pic, 
comet Hale-Bopp, and the zodiacal light.
\label{wdspec_sil}}
\end{figure}

\clearpage

\begin{figure}[th]
\epsscale{1}
\plotone{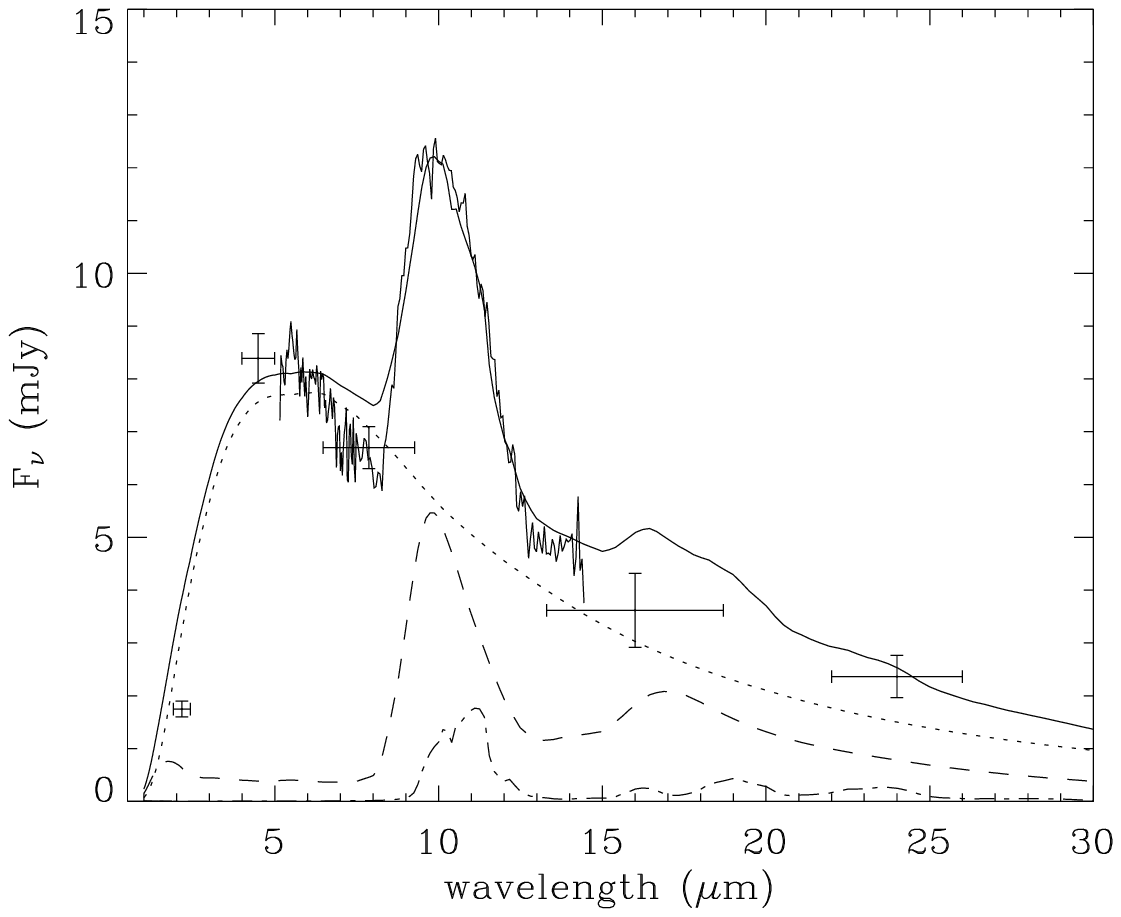}
\caption[f3.eps]{The infrared excess of G29-38 is compared to
the best-fit model (smooth solid line) consisting of three compositions: 
amorphous carbon (dotted line), amorphous olivine (long-dashed line),
and forsterite (short-long dashed line).
\label{fplot}}
\end{figure}
}


\begin{thebibliography}{} 

\bibitem[Alcock et al.(1986)]{alcock} Alcock, C., Fristrom, C. C., \& 
Siegelman, R. 1986, \apj, 302, 462
 
\bibitem[Alcock \& Illarionov(1980)]{alcock80} 
Alcock, C., \& Illarionov, A. 1980, ApJ, 235, 534

\bibitem[Bergeron et al.(1995a)]{postAGBage} Bergeron, P., Wesemael, F., 
\& Beauchamp, A. 1995, \pasp, 107, 1047

\bibitem[Bergeron et al.(1995b)]{bergeron95} Bergeron, P., Wesemael, F., Lamontagne, R.,
Fontaine, G., Saffer, R. A., \& Allard, N. F. 1995, \apj, 449, 258

\bibitem[Chary et al.(1999)]{chary} Chary, R., Zuckerman, B., \& 
Becklin, E. E. 1999, in {\it The Universe as Seen
by ISO}, eds. P. Cox \& M. F. Kessler, ESA-SP 427, p. 289

\bibitem[Crovisier et al.(1997)]{crovisier} Crovisier, J.,  Leech, K.,
Bockelee-Morvan, D., Brooke, T. Y., Hanner, M. S., Altieri, B., 
Keller, H. U., Lellouch, E. 1997, Science, 275, 1904

\bibitem[Debes \& Sigurdsson(2002)]{debes02} Debes, J. H., \& 
Sigurdsson, S. 2002, \apj, 572, 556

\bibitem[Dupuis et al.(1993)]{dupuis} 
Dupuis, J., Fontaine, G., \& Wesemael, F.  1993, \apjs, 87, 345

\bibitem[Fazio et al.(2004)]{fazio}
Fazio, G. G. et al. 2004, \apjs, 154, 10

\bibitem[Fixsen \& Dwek(2002)]{fixsendwek}
Fixsen, D. J., \& Dwek, E. 2002, ApJ, 578, 1009

\bibitem[Graham et al.(1990)]{graham}
Graham, J. R., Matthews, K., Neugebauer, G., \& Soifer, B. T. 1990, \apj, 357, 216

\bibitem[Green et al.(1986)]{green}
Green, R. F., Schmidt, M., and Liebert, J., 1986, \apjs, 61, 305

\bibitem[Greenstein \& Liebert(1990)]{greenstein} Greenstein, J. L,
\& Liebert, J. W. 1990, \apj, 360, 662

\bibitem[Gr\"un et al.(1985)]{gruen85}
Gr\"un, E., Zook, H. A., Fechtig, H., \& Giese, R. H. 1985. Icarus, 62, 244

\bibitem[Harrington \& Dahn(1980)]{usno}
Harrington, R. S., \& Dahn, C. C. \aj, 85, 454

\bibitem[Holberg et al.(2003)]{holberg} Holberg, J. B., Barstow, M. A.,
\& Burleigh, M. R. 2003, \apjs, 147, 145

\bibitem[Houck et al.(2004)]{houck}
Houck, J. R. et al. 2004, \apjs, 154, 18

\bibitem[Jura(2003)]{jura} Jura, M. 2003, \apjl, 584, L91

\bibitem[Kemper et al.(2004)]{kemper} Kemper, F., Vriend, W. J., \&
Tielens, A. G. G. M. 2004, ApJ, 6009, 826

\bibitem[Kleinman et al.(1998)]{kleinman} Kleinman, S. J. et al. 1998, ApJ, 495, 424

\bibitem[Koester et al.(1997)]{koester97}
	Koester, D., Provencal, J., \& Shipman, H. L. 1997, \aap, 230, L57

\bibitem[Kuchner et al.(1998)]{kuchner98} Kuchner, M. J., Koresko, C. D., \& Brown, M. E. 1998, \apjl, 508, L81

\bibitem[Kuchner \& Saeger(2005)]{kuchnersaeger} Kuchner, M. J., \& Saeger, S. 2005, submitted to ApJ

\bibitem[McCook \& Sion(1987)]{mccooksion} McCook, G. P. \& Sion, E. M. 1987, \apjs, 65, 603

\bibitem[Reach et al.(2003)]{reachzody} Reach, W. T., Morris, P., 
	Boulanger, F., \& Okumura, K. 2003, Icarus, 164, 384

\bibitem[Reach et al.(2004)]{reachcal} Reach, W. T., et al. 2005, PASP, in press

\bibitem[Rieke et al.(2004)]{rieke}
Rieke, G. H. et al. 2004, ApJS, 154, 25

\bibitem[Rieke et al.(2005)]{riekeAstar}
Rieke, G. H. et al. 2005, ApJ, 620, 1010

\bibitem[Sloan et al.(2003)]{sloan}
Sloan, G. C., Kraemer, K. E., Price, S. D., \& Shipman, R. F. 2003, ApJS, 147, 379

\bibitem[Sykes \& Greenberg(1986)]{sykes86}
Sykes, M. V., \& Greenberg, R. 1986, Icarus, 65, 51

\bibitem[Tokunaga et al.(1990)]{tokunaga}
Tokunaga, A. T., Becklin, E. E., \& Zukerman, B. 1990, \apjl, 358, L21

\bibitem[Werner et al.(2004)]{werner}
Werner, M. W. et al. 2004, ApJS, 154, 1

\bibitem[Weidemann(2000)]{weidemann} Weidemann, V. 2000, \aap, 363, 647

\bibitem[Zuckerman \& Becklin(1987)]{ZB87}
Zuckerman, B., \& Becklin, E. E. 1987, Nature, 330, 138

\bibitem[Zuckerman \& Reid(1998)]{zuckreid}
Zuckerman, B., \& Reid, I. N. 1998, \apjl, 505, L143

\bibitem[Zuckerman et al.(2003)]{zuckerman}
Zuckerman, B., Koester, D., Reid, I. N., \& H\"unsch, M. 2003, \apj, 596, 477

\end{thebibliography}
\end{document}